\documentclass[english,aps,preprint]{revtex4}
\usepackage[]{fontenc}
\usepackage[latin1]{inputenc}
\usepackage{graphicx}
\usepackage{amssymb}

\makeatletter

\usepackage{babel}
\makeatother
\begin{document}

\title{Comments on the black hole information problem}

\author{David A. Lowe}

\email{lowe@brown.edu}

\affiliation{Department of Physics, Brown University, Providence, RI 02912, USA}

\author{Lárus Thorlacius}

\email{lth@hi.is}

\affiliation{Science Institute, University of Iceland, Dunhaga 3, IS-107 Reykjavík,
Iceland}

\begin{abstract}
String theory provides numerous examples of duality between gravitational
theories and unitary gauge theories. To resolve the black hole information
paradox in this setting, it is necessary to better understand how
unitarity is implemented on the gravity side. We argue that unitarity
is restored by nonlocal effects whose initial magnitude is suppressed
by the exponential of the Bekenstein-Hawking entropy. Time-slicings
for which effective field theory is valid are obtained by demanding
the mutual back-reaction of quanta be small. The resulting bounds
imply that nonlocal effects do not lead to observable violations of
causality or conflict with the equivalence principle for infalling
observers, yet implement information retrieval for observers who stay
outside the black hole.
\end{abstract}
\maketitle

\section{introduction}

The black hole information paradox \cite{Hawking:1976ra} highlights
an incompatibility between locality in spacetime physics and unitarity
in quantum theories \cite{'tHooft:1991rb,Susskind:1993if}. The argument
for information loss is made in the context of a low-energy effective
theory on a semiclassical background spacetime, which is assumed to
be a conventional local quantum field theory. Unitarity in black hole
evolution, on the other hand, implies nonlocality in spacetime and
the nonlocal effects in question must act over macroscopic length
scales.

There is mounting evidence that the paradox is resolved in favor of
unitarity, coming for example from matrix theory \cite{Banks:1996vh},
the AdS/CFT correspondence \cite{Maldacena:1997re,Gubser:1998bc,Witten:1998qj},
and more indirectly from the microscopic computation of black hole
entropy \cite{Strominger:1996sh}. The AdS/CFT correspondence in particular
provides a framework where unitarity is manifest on the gauge theory
side, but since a semiclassical geometry is only recovered in a strong
coupling limit, the implementation of unitarity on the gravity side
remains a challenge. One must somehow reconcile the semiclassical
effective field theory with the apparent duplication of degrees of
freedom required for the information carried by matter falling into
the black hole to be returned in the Hawking radiation.

\begin{figure}
\includegraphics{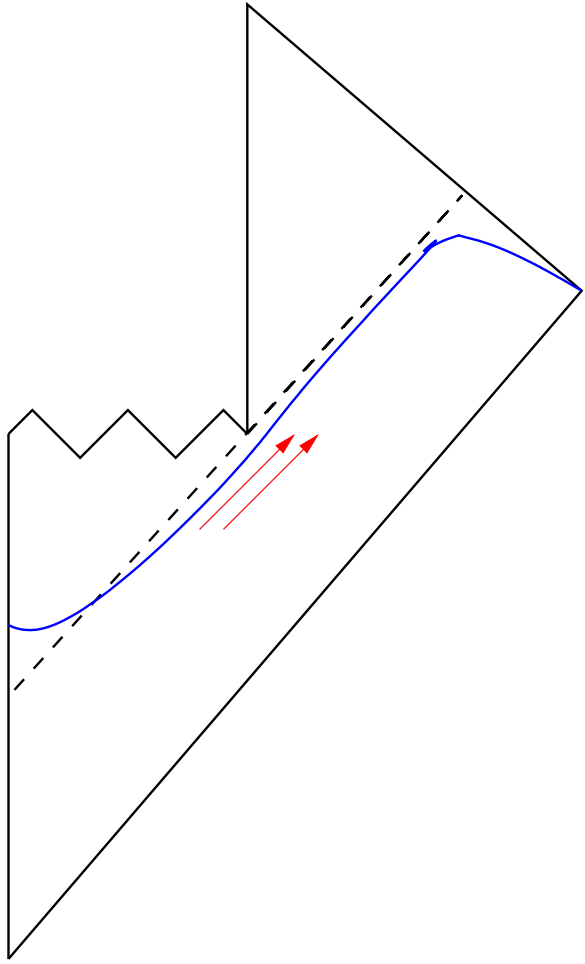}

\caption{\label{cap:figone} Timeslices can be constructed in the semiclassical
geometry that simultaneously intersect most of the Hawking radiation
and a freely falling observer who crosses the horizon early on.}
\end{figure}

These concerns only arise if one assumes that the evaporation process
is governed by a local effective field theory combining matter and
gravity. In particular, it is assumed that one can construct so called
nice time slices \cite{Wald:1993,Lowe:1995ac} that simultaneously
intersect most of the outgoing Hawking radiation and an infalling
observer having just passed inside the horizon in such a way that
both the outgoing radiation and the infalling observer have low energy
in the local frame of the slice, as indicated in figure \ref{cap:figone}.
These slices can be constructed to avoid regions of large spacetime
curvature, except near the black hole endpoint.

It was argued in \cite{Lowe:1995ac,Lowe:1995pu}, based on perturbative
string theory, that the effective field theory on such time slices
is in fact nonlocal on macroscopic length scales. In a similar vein,
\cite{Giddings:2001pt,Giddings:2004ud} argued that gravitational
back-reaction will undermine locality in the effective field theory
on nice time slices in a black hole geometry. We can turn these arguments
around to say that effective field theory is only to be applied on
restricted time slices. In the limit that we treat the near-horizon
region of a large black hole as Rindler space, we formulate the restriction
in terms of an upper bound on the relative boost of any two observers,
that are at rest with respect to different parts of a time slice.
When the restriction due to the boost bound is taken into account,
we find that observers crossing the horizon of a black hole will see
local and causal physics to within the finite accuracy of their measuring
apparatus until they approach the singularity, while outside observers
can see complete retrieval of information.

\section{The paradox\label{sec:The-paradox}}

We begin with a sharp statement of the black hole information paradox.
Consider a pair of spin $1/2$ particles in a singlet state\begin{equation}
\frac{{1}}{\sqrt{{2}}}\left(|\uparrow_{1}\rangle|\downarrow_{2}\rangle-|\downarrow_{1}\rangle|\uparrow_{2}\rangle\right),\label{eq:singlet}\end{equation}
where $|\uparrow\rangle,~|\downarrow\rangle$ refer to eigenvectors
with spin oriented along the $z$-axis. Spin \#2 is kept far outside
the black hole in the custody of observer $\mathcal{O}_{far}$, while
spin \#1 is sent, along with a free-falling observer $\mathcal{O}_{free}$,
into the black hole. Another observer $\mathcal{O}_{accel}$, hovers
outside the black hole and measures the Hawking radiation coming off.
Far away from the black hole $\mathcal{O}_{far}$ conducts a measurement
on spin \#2, which consists of either measuring the spin along the
$z$-axis or the $x$-axis. If the other observers can combine their
results to determine which of these possibilities has taken place,
then information has been acausally transmitted over an arbitrarily
long distance.

Suppose the Hawking radiation eventually carries all the information
about the internal state of the black hole. Then $\mathcal{O}_{accel}$
can conduct a measurement of the radiation to effectively determine
the component of spin \#1 along the $z$-axis. If $\mathcal{O}_{far}$
measures spin \#2 up along the $z$-axis, the indirect measurement
carried out by $\mathcal{O}_{accel}$ will give spin down. If $\mathcal{O}_{far}$
instead measures spin \#2 along the $x$-axis there will be no correlation
between the results of the two spin measurements. So far no actual
information has been acausally propagated, since the probability for
each observer to measure spin up or down along any axis is 50\%. Therefore
$\mathcal{O}_{accel}$ cannot know in which direction $\mathcal{O}_{far}$
conducted the measurement. Only if they were to later come into causal
contact and compare notes, would they notice the nonlocal EPR correlation
between their spin measurements.

Now suppose $\mathcal{O}_{free}$ makes a measurement on spin \#1
along the $z$-axis. Since $\mathcal{O}_{free}$ is at spacelike separations
from $\mathcal{O}_{accel}$ their measurements should commute with
each other if local quantum field theory is valid. Therefore $\mathcal{O}_{free}$
will measure spin up or down with 50\% probability. Again no information
has acausally propagated, despite the nonlocal EPR correlations between
spins \#1 and \#2.

Finally, $\mathcal{O}_{accel}$ enters the black hole, and compares
notes with $\mathcal{O}_{free}$. If they disagree on the outcome
of their experiments, they know with certainty that $\mathcal{O}_{far}$
must have conducted a measurement along the $x$-axis. Thus with some
finite probability, acausal communication has occurred between $\mathcal{O}_{far}$
and our intrepid observers inside the black hole. In essence, the
black hole operates as a quantum information cloning machine, and
such machines allow EPR correlations to be turned into EPR phones.
This experiment provides us with a clear statement of the information
paradox.

Due to the extreme redshifts involved, it turns out to be very difficult
for the observer $\mathcal{O}_{accel}$ to receive any signal from
observer $\mathcal{O}_{free}$ before crashing into the singularity.
In \cite{Susskind:1993mu} it was argued that $\mathcal{O}_{free}$
would have to employ an enormous frequency for the transmission and
the amount of energy involved would then drastically alter the background
geometry. The argument is made in two steps. First one needs an estimate
of the time $\mathcal{O}_{accel}$ has to wait outside the black hole
while the indirect measurement of spin \#1 is carried out. In the
second step one bounds the available time for observer $\mathcal{O}_{free}$
to perform the direct measurement on spin \#1 and transmit the result
for $\mathcal{O}_{accel}$ to receive before hitting the singularity. 

The 3+1-dimensional estimates of \cite{Susskind:1993mu} are easily
generalized to arbitrary dimensions. The relevant facts about a $D$-dimensional
black hole geometry are summarized in appendix \ref{sec:Black-hole-geometry}.
If we assume the Hawking radiation and black hole are in a random
pure state, as should be true if interactions effectively thermalize
the black hole degrees of freedom, then we may apply the results of
Page \cite{Page:1993df,Page:1993wv} (see appendix \ref{sec:Average-entropy-of}
for a brief review) to compute the typical amount of information contained
in the Hawking radiation. The information outflow is extremely small,
of order $e^{-S_{bh}}$ where $S_{bh}$ is the Bekenstein-Hawking
entropy, until one half-life of the black hole has passed, at a Schwarzschild
time $t\sim r_{s}S_{bh}$ where $r_{s}$ is the Schwarzschild radius,
after which the information contained in the Hawking radiation increases
linearly with the thermal entropy of the radiation. The fact that
$\mathcal{O}_{accel}$ has to wait such a long time before making
the plunge into the black hole leaves $\mathcal{O}_{free}$ with very
little time (of order $e^{-a\, S_{bh}}$, with $a$ an order one constant,
according to the estimate at the end of appendix\ref{sec:Black-hole-geometry})
to perform the direct spin measurement and send off the result. Accordingly
the signal would have to be encoded in a wave with frequency of order
$e^{a\, S}$, which in turn requires more energy than is available
to observers in this geometry.

\section{Unitary quantum mechanical framework\label{sec:Unitary-quantum-mechanical}}

Let us now rerun the gedanken experiment, in the type of quantum gravity
framework that AdS/CFT provides. Our key assumptions are: 

\begin{enumerate}
\item black hole evaporation is unitary, with all the information being
carried off by the Hawking radiation; 
\item a quantum mechanical description of the evaporation is valid.
\end{enumerate}
We have in mind black holes whose Schwarzschild radius is large compared
to any microscopic length scale but small compared to the characteristic
AdS length scale. In this case, asymptotically AdS boundary conditions
do not play a crucial role in our discussion, and for simplicity we
will draw Penrose diagrams with asymptotically flat boundary conditions.

Suppose $\mathcal{O}_{free}$ conducts a measurement of spin \#1,
with result spin up. Up to the subtleties described below, having
to do with the finite size of the black hole, this will project the
state into\begin{equation}
|\uparrow_{1}\rangle|\downarrow_{2}\rangle~.\label{eq:newstate}\end{equation}
 Further measurement of this state by $\mathcal{O}_{accel}$ will
necessarily lead to the same answer. Likewise $\mathcal{O}_{accel}$
and $\mathcal{O}_{free}$ will agree on the results of their measurements
regardless of the order with which they are conducted in some given
time slicing. This can only happen if the measurement operator from
the effective field theory viewpoint is nonlocal, or equivalently
the degrees of freedom in the effective field theory are nonlocal.
If $\mathcal{O}_{accel}$ and $\mathcal{O}_{free}$ are always in
agreement, then no information is propagated acausally over arbitrary
distance from $\mathcal{O}_{far}$. Likewise, no acausal propagation
of information is observed by those who stay outside the black hole.
Another concern is that observers inside and outside the black hole,
such as $\mathcal{O}_{free}$ and $\mathcal{O}_{accel}$, are able
to communicate. We reviewed the argument of \cite{Susskind:1993mu}
that this cannot happen in the previous section and the following
is another variant.

We find it useful to couch the discussion in the many worlds interpretation
of quantum mechanics, where one gives up the projection postulate
of the Copenhagen interpretation \cite{Zurek:1991vd,Tegmark:1997me}.
From this viewpoint the wavefunction always evolves deterministically
according to the linear Schrodinger equation. This is important for
the case at hand, because the outside observer conducts measurements
on the entire black hole state, which includes $\mathcal{O}_{free}$.
It is therefore appropriate to think of $\mathcal{O}_{free}$ as a
mesoscopic quantum subsystem, whose observations do not project the
quantum state. Rather, different mesoscopic states of $\mathcal{O}_{free}$
will be correlated with different values of the spin state. To the
extent that the dynamics experienced by $\mathcal{O}_{free}$ is local,
these different mesoscopic states will effectively decohere into quasi-classical
position eigenstates due to interactions with their environment \cite{Zurek:1991vd},
so may be interpreted as measurements by $\mathcal{O}_{free}$.

Furthermore the intrinsic accuracy of measurements carried out by
observers inside a black hole is limited, since the maximum size apparatus
they can have access to is bounded by a subsystem with Hilbert space
dimension $e^{S_{bh}}$. For example, suppose we have $N$ identical
spins, with $2^{N}=e^{S_{bh}}$. Then the intrinsic accuracy of a
spin measurement is proportional to \begin{equation}
\sigma_{spin}\sim\frac{1}{\sqrt{N}}\sim\frac{1}{\sqrt{S_{bh}}}\,.\label{eq:accuracy}\end{equation}
More generally, if $\mathcal{O}_{free}$ only has a finite time to
conduct a measurement, we can assume the maximum size of the apparatus
satisfies a holographic bound for the relevant causal diamond\begin{equation}
\sigma_{spin}\sim\frac{1}{\sqrt{A}}\label{eq:holobound}\end{equation}
where $A$ is the area of the relevant light-sheets \cite{Bousso:1999xy}.

As before, we invoke the results of Page \cite{Page:1993df,Page:1993wv}
to conclude that observer $\mathcal{O}_{accel}$ must wait for $t\sim r_{s}S_{bh}$
before useful information can be extracted from the Hawking radiation.
This means for $\mathcal{O}_{free}$ to communicate results to $\mathcal{O}_{accel}$
the signal must be sent in a small trans-Planckian time\begin{equation}
t\sim e^{-aS_{bh}},\label{eq:tptime}\end{equation}
with $a$ is a number of order one, else $\mathcal{O}_{accel}$ hits
the singularity before receiving the signal. This implies the intrinsic
accuracy in $\mathcal{O}_{free}$ measurements (\ref{eq:holobound})
is necessarily only of order one. Hence no useful information can
be transmitted to $\mathcal{O}_{accel}$.

We can nevertheless ask whether observations on the Hawking radiation
can affect the inside observer $\mathcal{O}_{free}$. Again it is
helpful to keep in mind the many worlds interpretation of quantum
mechanics. As the Hawking radiation moves away from the black hole,
it will be an increasingly good approximation to treat its interactions
as local. In the AdS/CFT correspondence the mapping to field theory
degrees of freedom is best understood as the boundary is approached
where it becomes local. Conversely, we expect the inside observer
whose state is encoded in this Hawking radiation will experience nonlocal
interactions. An outside observer can in principle employ arbitrarily
large measuring apparatus, so for all practical purposes can project
the black hole state into a position eigenstate of the Hawking radiation.
Since the inside observer's state is itself projected, it appears
that observation of the Hawking radiation in fact burns up the inside
observer!

It is possible to construct so called nice time slices which intersect
$\mathcal{O}_{free}$ shortly after crossing the horizon, and also
intersect the bulk of the Hawking radiation near the endpoint of the
evaporation. Such a time slice is indicated in figure \ref{cap:figone}
and an explicit construction is given in \cite{Lowe:1995ac}. If we
apply the decoherence story on such a time slice, the local coupling
of the Hawking radiation to environmental degrees of freedom far from
the black hole is already sufficient to burn up the freely falling
observer as soon as they cross the horizon, without any explicit observation
needed. While this scenario gives a consistent description for observers
far from the black hole, the conclusion is distinctly undemocratic.
By the equivalence principle an observer entering a large black hole
in free fall is expected to experience physics as usual until the
region of strong curvature near the singularity.

\section{Validity of effective field theory}

We were led to the preceding conclusion by assuming the validity of
a local effective field theory on the nice time slices. While this
assumption is central to the black hole information paradox, it is
quite subtle. The nice time slices are locally well behaved but from
a global viewpoint they involve very large numbers. The integrated
extrinsic curvature along each slice is enormous as can be seen, for
example, by parallel transporting a time-like unit normal vector from
the region of the outgoing Hawking radiation into the black hole along
the radial direction. The resulting vector is related by an enormous
local Lorentz boost to a timelike unit normal on this part of the
slice. For a large enough black hole we can apply the Rindler space
approximation to the near horizon region,\begin{equation}
r-r_{s}\ll r_{s}\,,\label{eq:rregion}\end{equation}
and simplify the discussion by neglecting the local spacetime curvature.
A nice time slice passing through the Rindler region contains observers
with large relative boosts from the point of view of the Minkowski
time of the region. A minimal condition for the validity of effective
field theory is obtained by requiring that a quantum wavepacket, with
energy close to some fixed proper cutoff scale $\Lambda$, say 1 TeV,
and localized near some arbitrary point on the time-slice, not produce
a large gravitational back-reaction on the semi-classical geometry,
as seen by another wavepacket at some well-separated point \cite{Giddings:2001pt,Giddings:2004ud}. 

In the Rindler space approximation, one can expect back-reaction to
be large when the conditions of the hoop conjecture are satisfied
\cite{ThorneK.S.1972}. Namely that horizons form when and only when
a mass $M$ can be surrounded by a hoop of circumference $4\pi GM$
in any direction. Because one can neglect spacetime curvature in the
Rindler space limit, notions of circumference and mass are well-defined. 

One criticism that has often been raised against claims of large effects
due to large relative boost is that the wavepackets in question are
receding from each other rapidly. When these wavepackets are traced
backward in time, they converge and at first sight it seems there
would have been a large interaction effect. One has similar issues
in cosmology, such as the trans-Planckian problem \cite{Martin:2000xs}.
Usually one takes the viewpoint that modes necessarily started out
in the adiabatic vacuum state, where no large effects are present.
At much later times, as the cosmological background expands, high
frequency modes red-shift down to accessible scales. In this case,
populating these modes at late-times need not produce any large effect.
In this situation, one can find time-slicings (for example, global
time in de Sitter space) where the conditions of the hoop conjecture
are only satisfied for wavepackets subject to the usual Jeans instability
on sub-horizon size scales, but not for localized modes separated
by super-horizon size distances. In the black hole case, the difference
is that although the two wavepackets are rapidly receding, nice slices
that avoid the singularity allow for wavepackets at small enough separations
for the hoop conjecture to apply. 

Therefore demanding that the influence of a particle at late-time
on the outside not dramatically change the black hole geometry, leads
to an upper bound on the allowed relative boost between observers
on a given time slice, \begin{equation}
\Lambda\gamma_{max}\ll M\,,\label{eq:bbound}\end{equation}
 where $M$ is the black hole mass %
\footnote{It is interesting to note that versions of dS/CFT using quantum deformations
of the conformal group provide concrete models that realize boost
bounds, already at the level of the free theory \cite{Guijosa:2003ze,Lowe:2004nw,Lowe:2005de}.%
}. Our interpretation of this type of the back-reaction bound differs
somewhat from that of \cite{Giddings:2001pt,Giddings:2004ud}. We
employ the back-reaction bound strictly as a condition for the validity
of an effective semiclassical description, allowing for the existence
of (small) nonlocal effects even when the bound is satisfied.

The bound (\ref{eq:bbound}) is certainly violated on the nice time
slices discussed above, where a significant fraction of the Hawking
radiation has been emitted, and therefore any conclusions drawn from
local effective field theory on such slices should be questioned.
In fact, the nonlocal condition on time slices implied by the boost
bound restricts the applicability of local effective field theory
rather severely in the black hole context. If a time slice intersects
both $\mathcal{O}_{free}$ and $\mathcal{O}_{accel}$ , and we take
$\mathcal{O}_{free}$ to be at Kruskal time $V=1$, then the Kruskal
time of $\mathcal{O}_{accel}$ must satisfy\begin{equation}
V=e^{v/r_{s}}\lesssim\gamma_{max}\,,\label{eq:gammamax}\end{equation}
in order to respect the boost bound on the time slice in question.
The maximum accessible Schwarzschild time measured by $\mathcal{O}_{accel}$
before the breakdown of local effective field theory is therefore
\begin{equation}
t_{max}\sim r_{s}\log\frac{M}{\Lambda}\,,\label{eq:tmax}\end{equation}
This is also roughly the time when the string theory calculation of
\cite{Lowe:1995pu} begins to depart from the corresponding field
theory calculation. Other arguments \cite{Schoutens:1993hu,Susskind:1993aa,Giddings:2004ud}
also lead to this time scale.

\begin{figure}
\includegraphics{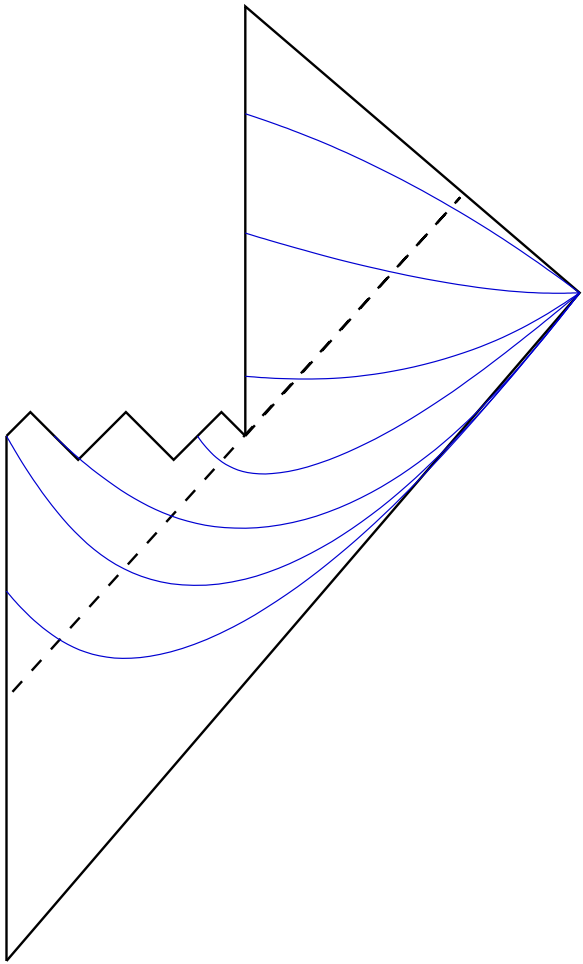}

\caption{\label{cap:A-set-of}A set of timeslices satisfying the back-reaction
bound, and for which an observer hovering close to the horizon remains
at low energy.}
\end{figure}

Since the time (\ref{eq:tmax}) is much smaller than the black hole
lifetime, $\tau\sim r_{s}S_{bh}$, then according to Page's arguments
the information contained in the Hawking radiation at this point is
only of order\begin{equation}
I\sim e^{-S_{bh}}\,,\label{eq:infocon}\end{equation}
where $S_{bh}$ is the initial Bekenstein-Hawking entropy of the black
hole. 

The Hawking effect was originally derived using local field theory
in a black hole background \cite{Hawking:1974sw}, so one might worry
that the breakdown of local effective field theory interferes with
the Hawking emission process itself. The timescale for the breakdown,
given in equation (\ref{eq:tmax}), is in fact also the time when
the first Hawking photons are emitted from the black hole. However,
by our estimate in equation (\ref{eq:infocon}), the nonlocal contribution
to the amplitude of any given Hawking photon is of order $e^{-S_{bh}}$
so the effect on the average Hawking flux will be negligible.

We note that information retrieval is non-perturbative in $1/N$ in
the AdS/CFT context. Let us apply our estimate to a black hole in
$AdS{}_{5}\times S_{5}\,.$ The relationship between the AdS length
scale and parameters of the gauge theory is given by\begin{equation}
R^{4}=4\pi g_{YM}^{2}N\,,\label{eq:adslength}\end{equation}
and for a semiclassical geometry we take $g_{YM}\rightarrow0$ and
$N\rightarrow\infty$ in such a way that $R\gg1$ is fixed. A physical
length on the gravity side, such as the Schwarzschild radius of a
black hole, scales the same way as $R$ with $N$, that is \begin{equation}
r_{s}\sim N^{1/4}\,.\label{eq:rswithn}\end{equation}
We are considering black holes with $r_{s}\ll R$, which are localized
in the ten-dimensional geometry. The black hole entropy (\ref{eq:bhentropy})
for $D=10$ scales as \begin{equation}
S_{bh}\sim r_{s}^{8}\sim N^{2}\,,\label{eq:sbhwithn}\end{equation}
and it follows that $e^{-S_{bh}}$ amounts to a non-perturbative effect
in $1/N$.

The information content (\ref{eq:infocon}) is a measure of how much
observations of the Hawking radiation perturb the quantum state inside
the black hole. From the point of view of the fundamental theory,
these really are the same degrees of freedom, so we will use (\ref{eq:infocon})
to estimate a lower bound on the size of non-local interactions in
the effective field theory. We expect these non-local interactions
can be expressed in a form local in time (\emph{c.f.} the light-cone
string field theory calculations of \cite{Lowe:1993ps,Lowe:1994ns,Lowe:1995ac})
since the underlying fundamental description, such as the conformal
field theory in the case of the AdS/CFT correspondence, is local in
time. Note that the asymptotic Schwarzschild time can be used to label
the slices. Hence the effective field theory will be nonlocal in space,
but otherwise will behave like a conventional quantum mechanical theory.

The magnitude of the nonlocal effects is of order $e^{-S_{bh}}$,
which is far below the intrinsic accuracy of $\mathcal{O}_{free}$'s
measurements (\ref{eq:accuracy}), and therefore unobservable. A timelike
observer in free fall will hit the singularity within a proper time
of order $r_{s}$ after passing through the horizon. The life expectancy
of the observer is in fact maximized by following a geodesic in free
fall. Any effort at accelerating away from the singularity only serves
to reduce the available proper time. The bound we have just established
holds along $\mathcal{O}_{free}$'s worldline all the way to the singularity.
Any time slice that satisfies the boost bound, and on which $\mathcal{O}_{free}$
is at low energy, can only intersect a small fraction of the outgoing
Hawking radiation. Thus only after $\mathcal{O}_{free}$ has hit the
singularity, can $\mathcal{O}_{accel}$ retrieve substantial information.
In this way the maximum boost bound explains why $\mathcal{O}_{free}$
has a reprieve from burn up until at the singularity, and is free
to observe local quantum mechanical evolution to within the accuracy
of accessible measurements. 

The outside observer $\mathcal{O}_{accel}$ can detect the tiny amounts
of information emitted in the Hawking radiation, since they have access
to arbitrarily large measuring apparatus, and in principle, can conduct
an arbitrarily large number of measurements on identically prepared
black hole states. Thus the non-local effects do act as a one-way
transmitter of acausal information, from the inside to the outside
of the black hole. 

Over the lifetime of the black hole, these small effects are sufficient
to emit all the information contained on the inside \cite{Page:1993wv}.
One might then worry that at late times, observation of this large
amount of information will bring us back to the burn-up scenario for
the inside observer. However if $\mathcal{O}_{free}$ is present on
a slice at late-times, we can infer they must have only recently fallen
across the horizon. In this case, we can apply Page's formula (\ref{eq:infocon})
to bound the information about $\mathcal{O}_{free}$ emitted in the
Hawking radiation, with $S_{bh}$ replaced by the entropy remaining
inside the horizon. This allows us to carry over the above estimates
of the influence of measurement on the inside observer, to conclude
such effects are undetectable on the inside.

This success comes at a price, however. The boost bound restricts
the use of effective field theory to foliations of spacetime, for
which the time slices terminate at the central singularity from a
relatively early time onwards, as indicated in figure \ref{cap:A-set-of}.
To make the effective field theory well-defined it will be necessary
to terminate each slice at relatively small spacetime curvature and
introduce boundary degrees of freedom to describe the high curvature
region. Presumably such a description can be obtained from the AdS/CFT
correspondence, but we do not know how to carry this out at present.
Therefore we content ourselves with placing bounds on the nonlocal
effects present in this effective field theory. Alternatively, we
could try to formulate an effective theory on a set of nice time slices
that avoid the strong coupling region but, since those slices run
afoul of the boost bound, unitarity would be implemented in such a
theory by nonlocal effects unsuppressed in magnitude.

\section{Discussion}

In quantum gravity, as defined by the AdS/CFT correspondence, black
hole evaporation is a unitary process generated by a conventional
Hamiltonian on the gauge theory side. The question then becomes how
this unitarity is implemented in an effective field theory in a black
hole background. The gauge theory correspondence provides a quantum
mechanical description of black hole evaporation which is local in
time, as measured by distant observers, but the evolution is not local
in the bulk spacetime. This is not surprising given that relation
between CFT operators and gravitational degrees of freedom is inherently
nonlocal, but that relation is of limited direct use due to the weak-to-strong
coupling nature of the duality. 

Above we estimated the minimal size of nonlocal effects in the effective
field theory for information to be returned in the Hawking radiation
and made some progress towards characterizing the nonlocality. Interestingly,
our order of magnitude estimate for nonlocal effects is comparable
to amplitudes involving topology change, computed using the Euclidean
path integral formulation of quantum gravity \cite{Maldacena:2001kr,Hawking:2005kf}. 

For an effective field theory picture to remain valid, the family
of allowed time slices must satisfy a back-reaction bound. Observation
of the Hawking radiation does disturb the state of observers inside
the black hole but, due to the bound, the inside observer does not
notice information extraction until near the singularity.

The arguments we have given in the present paper offer a more detailed
justification of the scenario described in \cite{Lowe:1999pk}, where
it was advocated based on the AdS/CFT correspondence that information
is extracted from inside observers only as they hit the singularity
but not at the horizon. A related idea was put forward in \cite{Horowitz:2003he}
where it was suggested that a unique semiclassical boundary condition
should be imposed at the singularity to enforce the extraction of
information. It is unclear to us, however, how to impose physical
boundary conditions on future spacelike surfaces. In our view the
same end is achieved by nonlocal dynamical effects as the singularity
is approached.

\begin{acknowledgments}
This research is supported in part by DOE grant DE-FG02-91ER40688-Task
A and by grants from the Science and Technology Policy Council of
Iceland and the University of Iceland Research Fund.
\end{acknowledgments}
\appendix

\section{Black hole geometry\label{sec:Black-hole-geometry}}

The metric of a Schwarzschild black hole in D-dimensional spacetime
is given by\begin{equation}
ds^{2}=-\left(1-\left(\frac{r_{s}}{r}\right)^{D-3}\right)dt^{2}+\left(1-\left(\frac{r_{s}}{r}\right)^{D-3}\right)^{-1}dr^{2}+r^{2}d\Omega_{D-2}^{2}\,,\label{eq:metric}\end{equation}
where the Schwarzschild radius $r_{s}$ is determined by the product
of the D-dimensional Newton's constant and the black hole mass,\begin{equation}
r_{s}^{D-3}=\frac{8\Gamma\left(\frac{D-1}{2}\right)}{\left(D-2\right)\pi^{\frac{\left(D-3\right)}{2}}}\, G_{N}^{\left(D\right)}M\,.\label{eq:rs}\end{equation}
Some values of $D$ are more interesting than others from the AdS/CFT
point of view but it can be useful to have expressions involving arbitrary
$D$. Strictly speaking we should be working with a more complicated
metric, which reflects the asymptotic AdS nature of the full geometry,
but it turns out the Schwarzschild metric (\ref{eq:metric}) is sufficient
for our purposes. We are considering black holes with $r_{s}\ll R$,
where $R$ is the AdS length scale, and in this case (\ref{eq:metric})
provides a good enough approximation in the region of the black hole
spacetime that is of interest to us.

The entropy of the black hole is given by $1/4$ of the area of the
event horizon,\begin{equation}
S=\frac{\pi^{\frac{\left(D-1\right)}{2}}}{2\Gamma\left(\frac{\left(D-1\right)}{2}\right)}\, r_{s}^{D-2}\,.\label{eq:bhentropy}\end{equation}
The Hawking temperature of the black hole is most easily obtained
by considering the Euclidean continuation $t\rightarrow-i\tau$ of
the metric (\ref{eq:metric}) and requiring the horizon to be smooth,\begin{equation}
T_{H}=\frac{\left(D-3\right)}{4\pi r_{s}}\,.\label{eq:hawkingtemp}\end{equation}
The lifetime $\tau$ of the black hole can be estimated from the Stefan-Boltzmann
law,\begin{equation}
\frac{dM}{dt}\sim-AT_{H}^{D}\sim-M^{-\frac{2}{D-3}}\,,\label{eq:stefanboltz}\end{equation}
from which it follows that\begin{equation}
\tau\sim M^{\frac{D-1}{D-3}}\sim r_{s}^{D-1}\,.\label{eq:bhlifetime}\end{equation}

The Schwarzschild coordinate system breaks down at the event horizon
$r=r_{s}$. For our purposes it is more convenient to work with the
D-dimensional analog of Kruskal-Szekeres coordinates. We first introduce
a tortoise coordinate $r_{*}$ via\begin{equation}
\frac{dr_{*}}{dr}=\left(1-\left(\frac{r_{s}}{r}\right)^{D-3}\right)^{-1}\,.\label{eq:tortoise}\end{equation}
For $D=4$ this integrates to $r_{*}=r+2M\log\left[\frac{r}{2M}-1\right]$
while for $D>4$ we obtain\begin{equation}
r_{*}=r+\frac{r_{s}}{D-3}\log\left[\frac{r}{r_{s}}-1\right]+F\left(r\right)\,,\label{eq:tort}\end{equation}
where $F\left(r\right)$ is finite as $r\rightarrow r_{s}$ . The
detailed form of $F(r)$ can be figured out but will not be needed
here. The next step is to define null coordinates, $v=t+r_{*}$ and
$u=t-r_{*}$, and finally these coordinates are exponentiated to obtain
\begin{equation}
V=\exp\left[\frac{\left(D-3\right)}{2r_{s}}\, v\right]\,,\qquad U=-\exp\left[-\frac{\left(D-3\right)}{2r_{s}}\, u\right]\,.\label{eq:kruskal}\end{equation}
The metric is non-singular at the event horizon in the $(U,V)$ coordinates,

\begin{equation}
ds^{2}=-\frac{4r_{s}^{2}}{\left(D-3\right)^{2}}\left(\frac{r_{s}}{r}+\left(\frac{r_{s}}{r}\right)^{2}\ldots+\left(\frac{r_{s}}{r}\right)^{D-3}\right)\exp\left[-\frac{\left(D-3\right)}{r_{s}}\left(r+F(r)\right)\right]dU\, dV\,.\label{eq:kruskalmetric}\end{equation}

\begin{figure}
\includegraphics{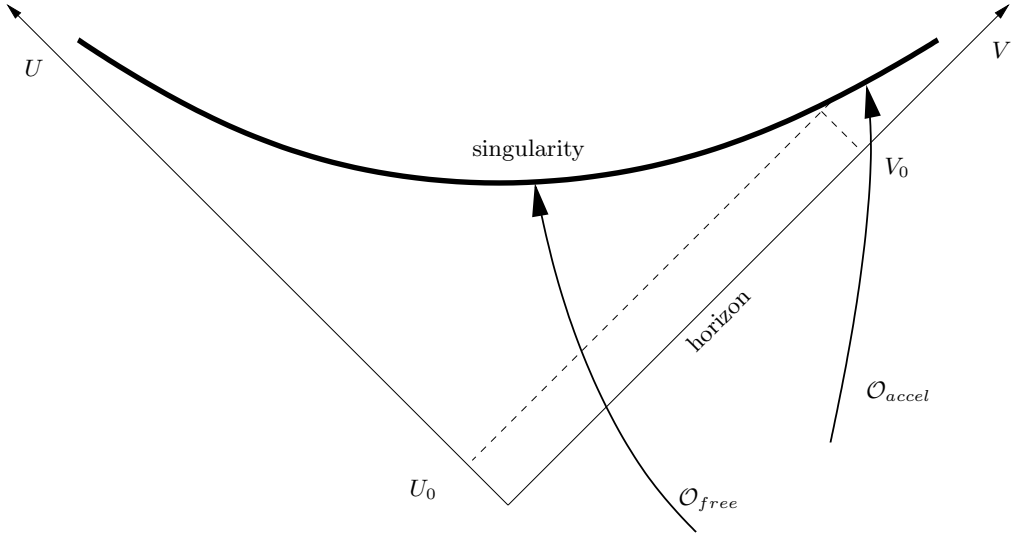}

\caption{\label{cap:kruskal} Signal propagation from ${\cal O}_{free}$ to
${\cal O}_{accel}$ is restricted by the black hole geometry.}
\end{figure}

We now have all the ingredients to make an estimate of the timescales
involved in the gedanken experiment described in section \ref{sec:The-paradox}.
The curvature singularity at $r=0$ is located on a curve of constant
$UV$ in the Kruskal coordinate system. Observer $\mathcal{O}_{accel}$
waits outside for half the black hole lifetime $t_{1/2}\sim r_{s}^{D-1}$
and enters the black hole at retarded Kruskal time $V=V_{0}\sim\exp(a'\, r_{s}^{D-2})\sim\exp(a\, S_{bh})\,,$
where $a'$ and $a$ are constants of $O(1)$. At this value of $V$
the singularity is at $U=U_{0}\sim1/V_{0}\sim\exp(-a\, S_{bh})\,,$
and the worldline of $\mathcal{O}_{accel}$ will inevitably run into
the singularity at an even smaller value of $U\,,$ as can be seen
in figure \ref{cap:kruskal}. 

Observer $\mathcal{O}_{free}$ is to perform the direct measurement
of spin \#1 after entering the black hole and communicate the result
to $\mathcal{O}_{accel}$ by a null signal. This task must be completed
while $\mathcal{O}_{free}$ is still in the causal past of $\mathcal{O}_{accel}$
(see figure \ref{cap:kruskal}). It must in other words be carried
out on the section of $\mathcal{O}_{free}$'s worldline that lies
within the narrow range $0<U<U_{0}\,.$ For a large black hole this
part of the worldline is in the Rindler region, where spacetime is
flat to a good approximation, and the available proper time for $\mathcal{O}_{free}$
to make the measurement and transmit the result is easily seen to
be of order $U_{0}\sim\exp(-a\, S_{bh})\,.$

It should be noted that there exist black brane geometries in asymptotically
AdS spacetime where the timescales involved in estimates of this type
appear to allow $\mathcal{O}_{accel}$ to receive messages from $\mathcal{O}_{free}$
well before $\mathcal{O}_{accel}$ runs afoul of the singularity.
In all such geometries that we know of, however, the entropy is lower
than that of black holes of the same mass density. Such a brane is
therefore subject to a Gregory-Laflamme type instability \cite{Gregory:1993vy,Reall:2001ag}
to decay into a collection of black holes on a relatively short timescale
and our estimates should instead be applied to the resulting black
holes.

\section{Average entropy of a subsystem \label{sec:Average-entropy-of}}

Consider a quantum system with a Hilbert space of dimension $mn$,
which is in a random pure state. Then the average entropy of a subsystem
of dimension $m<n$ is given by \begin{equation}
S_{m,n}=\sum_{k=n+1}^{mn}\frac{1}{k}-\frac{m-1}{2n}.\label{eq:sexact}\end{equation}
This result was conjectured by Page \cite{Page:1993df} and later
proved by Sen \cite{Sen:1996ph}. The information content of the subsystem
can be characterized by the difference between $S_{m,n}$ and the
maximum thermal entropy\begin{equation}
I_{m,n}=\log m-S_{m,n}\,.\label{eq:infomn}\end{equation}

The quantum system that we are interested in consists of an uncharged,
non-rotating black hole, along with all the Hawking radiation that
it has emitted at a given time, as measured by a distant observer
in the asymptotic rest frame of the black hole. We take as our subsystem
the train of Hawking radiation propagating away from the black hole
at a time when the remaining horizon area is more than half that of
the original black hole. In this case $n=e^{S_{bh}}$, with $S_{bh}$
the remaining black hole entropy, and we have \begin{equation}
n\geq m\gg1\,.\label{eq:nmggone}\end{equation}
In this limit the formula (\ref{eq:sexact}) becomes \cite{Page:1993df}\begin{equation}
S_{m,n}=\log m-\frac{m}{2n}+\mathcal{O}(\frac{1}{n^{2}})\,,\label{eq:approxentropy}\end{equation}
and the information content of the outgoing radiation is \begin{equation}
I_{m,n}=\frac{m}{2n}+\mathcal{O}(\frac{1}{n^{2}})\,.\label{eq:approxinfo}\end{equation}
The subsystem with $m<n$ carries less than one half a bit of information.
As described in \cite{Page:1993wv} once the radiation has carried
off half the initial entropy of the black hole, the information increases
linearly with $\log m$, the number of Hawking photons.

We are also interested in the information content of the Hawking radiation
at the onset of evaporation of the black hole. Consider for example
the case where the subsystem consists of the polarization degree of
freedom of a single Hawking photon only. For $m=2$, $n\gg m$ (\ref{eq:sexact})
reduces to\begin{equation}
S_{2,n}=\log2-\frac{3}{2n}+\mathcal{O}(\frac{1}{n^{2}})\,.\label{eq:earlyentropy}\end{equation}
The average information carried in the polarization of an early Hawking
photon is therefore exponentially suppressed,\begin{equation}
I_{2,n}\simeq\frac{3}{2n}\sim e^{-S_{bh}},\label{eq:earlyinfo}\end{equation}
 where $S_{bh}$ is the initial black hole entropy.

\bibliographystyle{brownphys}
\bibliography{comments}

\end{document}